\documentstyle[prl,aps,twocolumn,epsfig]{revtex}

\begin{document}
\twocolumn[\hsize\textwidth\columnwidth\hsize\csname @twocolumnfalse\endcsname

\title{Polarization rotation {\it via} a monoclinic phase in the piezoelectric 92\%PbZn$_{1/3}$Nb$_{2/3}$O$_3$-8\%PbTiO$_3$}

\author{B. Noheda\thanks{%
Corresponding author, e-mail: noheda@bnl.gov. On leave from Universidad Autonoma de Madrid}$^1$, D.E. Cox$^1$, G. Shirane$^1$, S-E. Park\thanks{%
Present address: Fraunhofer Technology Center, Hialeah, FL 33010}$^2$, L.E. Cross$^2$ and Z. Zhong$^3$}
\address{$^1$ Physics Department, Brookhaven National Laboratory, Upton, NY 11973}
\address{$^2$ Materials Research Laboratory, Pennsylvania State University, University Park, PA 16802}
\address{$^3$ National Synchrotron Light Source, Brookhaven National Laboratory, Upton, NY
11973}
\maketitle

\begin{abstract}
The origin of ultrahigh piezoelectricity in the relaxor ferroelectric 
PbZn$_{1/3}$Nb$_{2/3}$O$_3$-PbTiO$_3$ was studied with an electric field applied along 
the 001 direction. The zero-field rhombohedral R phase starts to follow 
the direct polarization path to tetragonal symmetry via an intermediate 
monoclinic M phase, but then jumps irreversibly to an alternate path 
involving a different type of monoclinic distortion. Details of the 
structure and domain configuration of this novel phase are described. This 
result suggests that there is a nearby R-M phase boundary as found in the 
Pb(Ti,Zr)O$_3$ system.
\end{abstract}

\vskip1pc]

\narrowtext
Recently two important advances have been made in searching for the origin of high piezo response 
in the perovskite oxides. One is a theoretical consideration\cite{Cohen,Garcia,Bellaiche} of the polarization rotation path 
under an electric field, 
emphasized by Fu and Cohen\cite{Cohen}. The second is the experimental discovery of a
monoclinic phase in Pb(Zr,Ti)O$_3$\cite{Noheda2,Guo} close to a "morphotropic" phase boundary (MPB), a nearly vertical
line between the rhombohedral and tetragonal phases. In this paper we demonstrate that these 
two seemingly disparate aspects are closely connected. 

Solid solutions of Pb(Zn$_{1/3}$Nb$_{2/3}$)O$_{3}$ containing a few
percent of PbTiO$_{3}$ (PZN-xPT) are relaxor ferroelectrics with ultrahigh
 piezoelectric responses an order of magnitude 
larger than those of conventional piezoelectric ceramics\cite{Kuwata1,Kuwata2,Park1}. 
They have a cubic perovskite-type structure at high temperatures, but undergo 
a diffuse ferroelectric phase transition at lower temperatures\cite{Kuwata1}. 
Materials in the ferroelectric region have structures with either rhombohedral or tetragonal symmetry, 
separated by a "morphotropic" phase boundary (MPB) at x$\simeq $ 10\%, similar to that 
of the well-known piezoelectric system PbZr$_{1-x}$Ti$_x$O$_3$ (PZT)\cite{Jaffe}. Extraordinarily high levels of electromechanical coupling 
and strain have been reported by Park and Shrout\cite{Park1} in rhombohedral crystals that are poled along 
 $\left[ 001\right] $, despite the fact that the polar
axis lies along $\left[ 111\right] $\cite{Kuwata2,Park2}. 

A typical strain-field loop for the rhombohedral composition PZN-8\%PT (close to the MPB) 
with an electric field applied along $\left[ 001\right] $ is plotted in Fig. 1 (top), after Durbin et al.\cite{Durbin1}.
 For this poling configuration, the strain behaves linearly below a certain threshold field, 
which is smaller for compositions closer to the MPB\cite{Park1,Park2}. This feature, 
together with the high strain levels obtained, makes these materials very promising for 
actuator applications. Above the threshold field, the strain shows a sharp jump, non-linear 
behavior and hysteresis. X-ray diffraction experiments performed on PZN-8\%PT 
crystals under an applied electric field \cite{Durbin1} have shown that the lattice parameter vs. field loop exactly reflects the strain behavior, confirming that the high macroscopic strain levels observed are due to the microscopic strain of the crystal lattice. The jump observed at the threshold field was attributed to a structural phase transition between the rhombohedral and 
the tetragonal phases. Very recently, it has been also reported
that the high-field tetragonal phase changes, after removal of the field, 
into a mixture of a dominant monoclinic phase with a remnant tetragonal phase\cite{Durbin2}. 

\begin{figure}[htb]
\epsfig{width=0.7\linewidth,figure=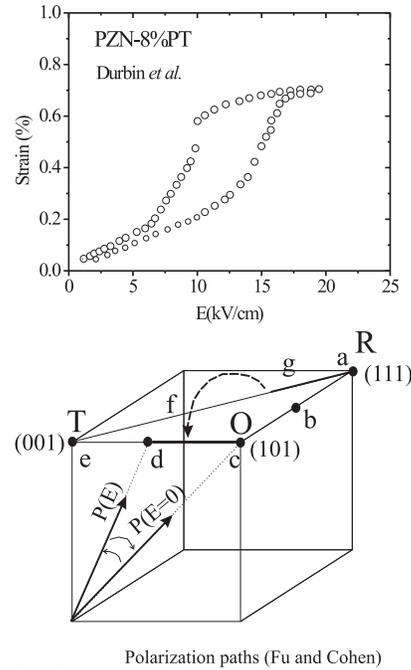}
\caption{Top: Strain-field loop for PZN-8\%PT (after Durbin et 
al.\protect\cite{Durbin1}). Bottom: Scheme of the two possible paths (R-T and R-O-T) for the polarization direction to change from $\left[ 111\right] $ in the rhombohedral (R) phase to $\left[ 001\right] $ in the tetragonal (T) phase as proposed by Fu and Cohen \protect\cite{Cohen}. The thickest lines represent the path followed by the polarization in PZN-8\%PT under the application of the electric field, as demonstrated by the present work. The solid arrows illustrate the orientation of the polarization in a particular domain with and without an electric field.}
\label{fig1}
\end{figure}

Our current study 
with high energy x-rays of 67 keV gives a very different picture. We show that the rhombohedral phase in the unpoled sample transforms into a purely monoclinic phase by the application of an electric field along the $\left[ 001\right] $ direction which, surprisingly,  remains monoclinic after the field is removed. This monoclinic phase, in which 
$b_m$ is directed along the pseudo-cubic $\left[ 010\right] $  axis, is 
not the same as that observed in PZT, in which $b_m$ lies along pseudocubic $\left[ 110\right] $ \cite{Noheda2}. The monoclinic phase in PZT represents the a-g-f-e path 
(see Fig. 1 bottom) proposed by Fu and Cohen \cite{Cohen} for the polarization to rotate between 
the rhombohedral and the tetragonal phases, while the monoclinic PZN-8\%PT phase represents the rotation path along a-c-d-e. 

Four crystals with composition 92\%PbZn$_{1/3}$Nb$_{2/3}$O$_{3}$ -8\%PbTiO$%
_{3} $ (PZN-8\%PT), grown by the high temperature
flux technique \cite{Park1,Park3}, were kindly provided by the authors of Refs. 
\onlinecite{Durbin1,Durbin2}. The crystals were oriented using a Laue back-reflection camera as described previously \cite{Park1}, and cut into 2x2x2 mm$^{3}$ cubes with their faces perpendicularto the $\langle $100$\rangle $ directions.

\begin{figure}[htb]
\epsfig{width=1.1 \linewidth,figure=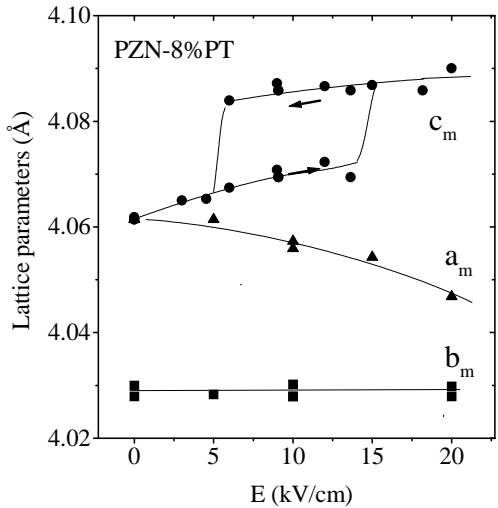}
\caption{Evolution of the lattice parameters, $a_m$,$b_m$ and $c_m$, of PZN-8\%PT with electric field} 
\label{fig2}
\end{figure}

Synchrotron x-ray diffraction measurements were initially performed at beamline X7A, at the National Synchrotron Light Source (NSLS), with x-rays of about 18 keV from a Si(111) double-crystal monochromator. Due to the high lead content, the samples are quite opaque at these energies, with a penetration depth of only about 1 $\mu $m, at the typical scattering angles used, and the measurements are therefore sensitive to inhomogeneous piezoelectric strains close to the surface (skin effect). In order to avoid this effect, high energy x-ray experiments were perfomed at 67 keV at the superconducting-Wiggler beamline X17B at the NSLS.

The 67 keV x-rays were obtained with a  Si (220) Laue-Bragg monochromator and the measurements were made on a four-circle Huber diffractometer equipped with a Si(220) analyzer, with an electric field applied to the crystal {\it in situ}. For this purpose thin wires were attached to the sample electrodes using a small drop of conducting epoxy, and a silicon grease with a high dielectrical breakdown was used to prevent arcing.
	
As-grown crystals of the composition under study are rhombohedral \cite{Kuwata1,Park1}(see Table I). However, when an electric field above a certain value is applied, the crystal symmetry changes and the rhombohedral phase can be recovered only either by annealing above the Curie temperature or by crushing the crystal. The initial rhombohedral state and the phase transition(s) undergone during the first field cycle are currently under investigation by Ohwada \cite{Ohwada}.

Once the rhombohedral phase is irreversibly transformed into the monoclinic one, the monoclinic lattice shows reversible changes as a function of the electric field, as shown in Fig. 2 (see also Table I). The monoclinic angle, $\beta = 90.15^{\circ }$, and the monoclinic axis, $b_m$, are field independent in the studied range between 0 and 20 kV/cm. $c_m$ increases with the electric field and shows a jump around 15 kV/cm and hysteresis, in agreement with Durbin et al. \cite{Durbin1}, while $a_m$ decreases as the field is increased and approaches the $c_m$ value in the E=0 limit \cite{Ref}.

The domain configuration observed in this experiment is illustrated in the representations 
of the reciprocal space shown in Fig. 3. \cite{Ref2}. Usually monoclinic symmetry leads to a very complicated domain configuration. However, once the field is applied, a much simpler situation preveals. The c-axis is now fixed along the field direction. As shown in the representation of the HK0 plane in Fig. 3a, there are only two {\it b} domains related by a $90 ^{\circ }$ rotation about the c-axis. Each of these {\it b} domains contains two {\it a} domains in which $a_m$ forms angles of either $\beta $ or $180^{\circ } -\beta $ respectively with $c_m$ (see Fig. 3b). The polarization vectors in each of the four domains, which rotate under the electric field within the $ac$ plane, form identical angles with the $\left[ 001\right] $ direction. It is evident from Fig. 3a that the domain twinning confers upon the lattice a pseudo-tetragonal symmetry, which might account for the previous assignment of tetragonal symmetry to this phase based on optical measurements\cite{Park3}. 

\begin{table}[tbp]

\caption{Lattice parameters, unit cell volume and symmetry of PZN-8\%PT as-grown, under a high field of 20 kV/cm, and after the field is removed. S denotes the crystal symmetry.}
\begin{tabular}{ccccccc}

& a (\AA ) & b (\AA ) & c (\AA ) & $\alpha $ /$\beta $ ($^{\circ })$ & V (\AA $^{3}$) & S \\ \hline
As-grown, E=0$^{\ast }$  & 4.053 & 4.053 & 4.053 & 89.90 & 66.58 & R \\ 
E=20kV/cm$^{\S }$ & 4.047 & 4.029 & 4.086 & 90.16 & 66.62 & M \\ 
After field, E=0$^{\ast }$$^{\S }$  & 4.061 & 4.030 & 4.061 & 90.15 & 66.46 & M \\ 
\hline
\end{tabular}

$^{\ast }${\small From powder diffraction.}
$^{\S }$ {\small From single crystal diffraction.}

\end{table}

Some examples of the x-ray peak profiles are shown in Fig. 4. The longitudinal scans over the (002) reflection at 0 kV/cm and at 20 kV/cm shown in the upper left of the figure illustrate the drastic change of the c-axis with the electric field (L= 2 corresponds to the $c_m=a_m$ value at E= 0). The very narrow mosaic at high fields is apparent in the upper-right of the figure, which shows a transverse scan (L scan) over the (0 2 0) Bragg peak at E= 20 kV/cm (K= 2 corresponds to the $c_m=a_m$ value at E= 0 so K= 2.015 means $b_m/a_m= 2/2.015$). The lower part of the figure is a contour map of the H0L zone around (200) after removal of the field, which shows the $b$ domain containing the two $a$ domains illustrated in Fig. 3a. and 3b. The symmetric splitting along the transverse direction (L scan), corresponds to the twin angle between the $a$ domains, so the L coordinate of 0.005 corresponds to a monoclinic angle $\beta $= $90.16 ^{\circ }$, that is found to be independent of the electric field. Although the (020) peak is very narrow, a small shoulder can be observed at positive L values (also present at 20 kV/cm, in the upper-right plot), which probably results from the crystal mosaic. 

\begin{figure}[h]
\epsfig{width=1.0 \linewidth,figure=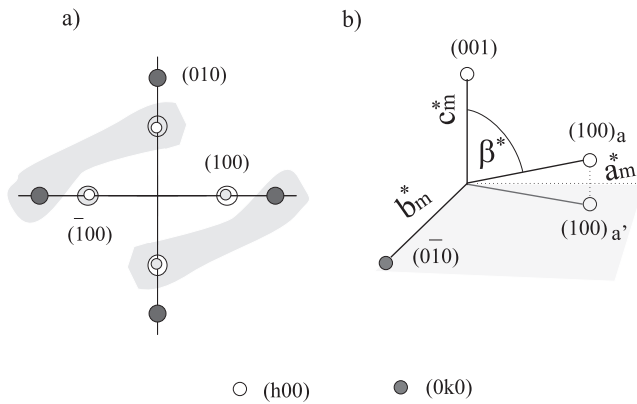}

\caption{(a) HK0 plane in reciprocal space for monoclinic PZN-8\%PT showing the two $b$ domains at $90^{\circ }$, shaded and unshaded respectively, in which H and K are interchanged with respect to each other. The double circles represent the two $a$ domains, illustrated in Fig. 3b. b) Scheme of the reciprocal unit cell of one of the $b$ domains showing the two $a$ domains, $a$ and $a'$, respectively.} 
\label{fig3}
\end{figure}

The low energy experiments at 18 keV essentially reproduce the results obtained by Durbin et al. cite{Durbin1,Durbin2}. The $c_m$ lattice parameter along the field direction $\left[ 001\right] $  exhibits the same strongly non-linear field dependence illustrated in Fig. 2. Measurements made with the scattering vector perpendicular to the field direction after the field was removed, show the same characteristic diffraction features  as recently reported protect\cite{Durbin2}, namely, a very broad peak on the low-angle side and a very sharp peak on the high-angle side, corresponding to H= 2.000 and K= 2.015, respectively. The broad peak changes with field while the sharp peak is field-independent, in agreement with our high-energy x-ray results described above. The markedly different widths of these two peaks led Durbin et al. to postulate the coexistence of two phases (monoclinic and tetragonal) at E= 0 kV/cm, with very different domain sizes. However, since the broadening of the (h00) peaks was not observed with high energy x-rays, it can be attributed to a "skin" effect.

Powder diffraction data collected from a sample prepared by crushing a small crystal piece 
after the application and removal of an electric field have confirmed that the monoclinic indexing of our single crystal experiments is correct. Lattice parameters $a_m$= $c_m$= 4.061 \AA, $b_m$= 4.030 \AA, and $\beta $= $90.15^{\circ }$ were extracted, in complete agreement with the single crystal data. In this case, however, the monoclinic phase was found to  coexist with a rhombohedral phase with $a_r$= 4.054 \AA, and $\alpha $ = $89.9^{\circ }$ (see Table I). 

\begin{figure}[h]
\epsfig{width=0.9 \linewidth,figure=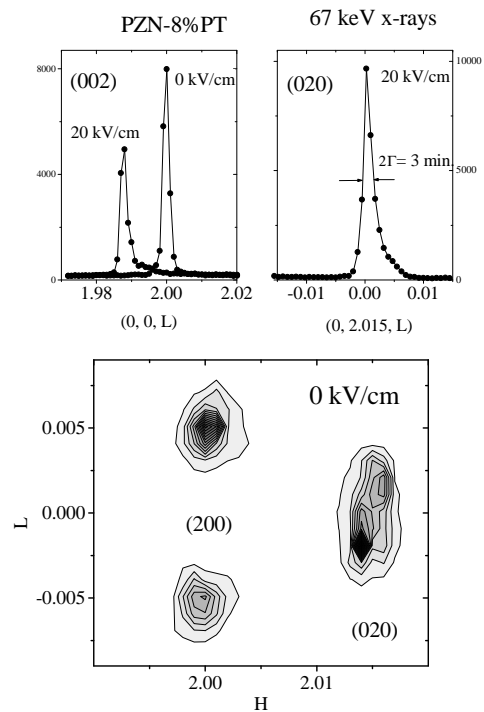}
\caption{Radial (L) scans over the (002) reflection at 20 kV/cm and at 0 kV/cm, after the field is removed
 (top left). The transverse (L) scan around (0, 2.015, 0) at 20 kV/cm is shown at the upper-right plot. At the bottom of the figure, an H0L mesh scan around the (200) Bragg reflection 
after the removal of the electric field. H= K= L= 2.0 have been defined to correspond 
to the $a_m$ value at E= 0 kV/cm.} 
\label{fig4}
\end{figure}

Comparison of our experimental results with the theoretical model of Fu and Cohen \protect\cite{Cohen} brings out several interesting features. These authors calculated the free energies along the polarization rotation paths between the rhombohedral R and tetragonal T points for BaTiO$_3$ (see Fig. 1), and concluded that the R-T path (a-g-f-e) is the most energetically-favorable one. PZN-8\%PT starts out at the R point in Fig. 1, since the as-grown crystal  is rhombohedral, but then changes irreversibly to the O-T path (c-d-e), with monoclinic symmetry. As mentioned above, this process of change between the rhombohedral and the monoclinic phases (dashed arrow in Fig. 1) is being studied elsewhere  in detail \cite{Ohwada}. When the field is decreased, the crystal remains on the O-T path. During this process $a_m$ and $c_m$ gradually approach each other until they eventually reach the same value at E= 0 kV/cm \protect\cite{Ref}. When the field is increased, $a_m$ approaches $b_m$ (see Fig.2), while following
 the O-T path, approaching the T point in Fig. 1. The polarization of each domain rotates within the (010) plane, and the polarization direction gets closer to the tetragonal polar axis, $\left[ 001\right] $, as the field is increased, as shown by the solid arrows in Fig. 1. However, the T point is not yet reached at the maximum electric fields (E= 20 kV/cm) used in the present study. Work is being carried out on other compositions, such as PZN-4.5\%PT and PZN-9\%PT, to determine the compositional range in which this behavior is observed. The major remaining question is why and how the sudden change at 15 kV/cm
 is reflected only in $c_m$. The jump observed in $c_m$ is very sensitive to the mechanical stress and can be completely eliminated if the sample is slightly stressed\cite{Viehland}. Work is in progress to address this point.

The electric field experiments reported by Guo et al. \protect\cite{Guo}, on rhombohedral PZT showed that, by poling at high temperatures, a monoclinic phase is also induced by the electric field  and retained after the field is removed. This monoclinic cell is such that $b_m$ lies along pseudo-cubic $\left[ 110\right] $  \protect\cite{Noheda2}, indicating that the polarization moves along the R-T (e-g-f-e) path when the field is applied. On the other hand, unpoled PZT has a monoclinic phase region from x= 0.46 to x= 0.52 at 20K \cite{Noheda3}. The existence of an irreversible rhombohedral-to-monoclinic phase transition induced by the application of an electric field in PZN-8\%PT, suggests that, as in the PZT case \cite{Guo}, a rhombohedral-to-monoclinic phase boundary exists close to x=8\% in the unpoled PZN-xPT phase diagram. This would clarify some of the apparent puzzles reported, such as the optical observation of symmetry lowering in PZN-9\%PT\protect\cite{Uesu,Fujishiro}. If the monoclinic phase is found to exist in unpoled PZN-9\%PT, then the phase diagram of PZN-PT becomes almost identical to that of PZT. Our current results suggest that the near degeneracy between the two monoclinic paths (R-T and R-O-T) might be responsible for the larger piezoelectric response in PZN-xPT, compared with that of PZT, in which only rotation is allowed. We expect that first-principles calculations may well clarify some of these important questions.

After submission of this letter, the very relevant work by D. Vanderbilt and M.H. Cohen 
\cite{Vanderbilt} came to our attention. By means of an eighth-order expansion of the Devonshire theory, these authors were able to predict three different monoclinic phases. Two of these, M$_A$ and M$_C$, correspond, respectively, to that previously observed in PZT and the one in PZN-8\%PT described above.

We would like to thank M. Durbin, Y. Fujii, P. Gehring, R. Guo, J. Hastings, K.
Hirota, K. Ohwada, D. Vanderbilt and T. Vogt for helpful discussions. Financial support by
the U.S. DOE under contract No. DE-AC02-98CH10886, by ONR Grant No.
 N00014-98-1-0527 and project MURI (N00014-96-1-1173) is also acknowledged.


\begin{references}
\bibitem{Cohen} H. Fu and  R.E. Cohen, Nature {\bf 403}, 281 (2000).
\bibitem{Garcia} A. Garcia and D. Vanderbilt, Appl. Phys. Lett. {\bf 72},2981 (1998). 
\bibitem{Bellaiche}  L. Bellaiche, A. Garcia, and D. Vanderbilt, Phys. Rev.
Lett. {\bf 84},5427 (2000).
\bibitem{Noheda2}  B. Noheda, J. A. Gonzalo, L.E. Cross, R. Guo, S-E. Park,
D.E. Cox, and G. Shirane. Phys. Rev. B {\bf 61},8687 (1999)
\bibitem{Guo}  R. Guo, L.E. Cross, S-E. Park, B. Noheda,D.E. Cox, and G.
Shirane, Phys. Rev. Lett. {\bf 84},5423 (2000)
\bibitem{Kuwata1}  J. Kuwata, K. Uchino, and S. Nomura, Ferroelectrics, {\bf 37},
579 (1981).

\bibitem{Kuwata2}  J. Kuwata, K. Uchino, and S. Nomura, Japan. J. of Applied
Phys. {\bf 21},1298 (1982).

\bibitem{Park1}  S-E. Park and T.R. Shrout, J. Appl. Phys. {\bf 82},1804
(1997).
\bibitem{Jaffe}  B. Jaffe, W.R. Cook, and H. Jaffe, Piezoelectric Ceramics,
Academic Press, London (1971).

\bibitem{Park2}  S-F. Liu, S-E. Park, T. R. Shrout, and L. E. Cross, J.
Appl. Phys. {\bf 85},2810 (1999).

\bibitem{Durbin1}  M.K. Durbin, E.W. Jacobs, J.C. Hicks, and S.-E. Park,
Appl. Phys. Lett. {\bf 74},2848 (1999).



\bibitem{Durbin2}  M.K. Durbin, J.C. Hicks, S-E. Park, and T.R. Shrout, J.
Appl. Phys., {\bf 87},8159 (2000)


\bibitem{Ohwada} K. Ohwada, (private communication).

\bibitem{Ref} In the limit at which $a_m = c_m$ and  $\beta > 90^{\circ }$, the crystal symmetry becomes orthorhombic, corresponding to the O point in Fig.1 bottom. 

\bibitem{Ref2} The real and the reciprocal lattice parameters are related by: $a_{m}^{\ast }$=$2\pi /(a_{m}\sin \beta ),$ $b_{m}^{\ast }$=$2\pi /b_{m},$ $%
c_{m}^{\ast }$=$2\pi /(c_{m}\sin \beta )$ and $\beta ^{\ast }=180^{\circ }-\beta $

\bibitem{Park3}  D-S. Paik, S-E. Park, S. Wada, S-F. Liu, and T. R. Shrout, J.
Appl. Phys. {\bf 85},1080 (1999).

\bibitem{Viehland} D. Viehland, J. Appl. Phys. {\bf 88}, 4794 (2000). 
\bibitem{Uesu} Y. Uesu, Y. Yamada, K. Fujishiro, H. Tazawa, S. Enokido, J.-M. Kiat, and B. Dkhil, 
Ferroelectrics {\bf 217},319 (1998)

\bibitem{Fujishiro} K. Fujishiro, R. Vlokh, Y. Uesu, Y. Yamada, J.-M. Kiat, B. Dkhil and Y. Yamashita, Jpn. J. Appl. Phys. {\bf 37},5246 (1998).


\bibitem{Noheda3} B. Noheda, D.E. Cox, G. Shirane, R, Guo, B. Jones, and L.E. Cross,  Phys. Rev. B. {\bf 63}, (to appear 01/01/2001).

\bibitem{Vanderbilt} D. Vanderbilt and M.H. Cohen, Phys. Rev. B. (in press), cond-mat/0009337.

\end{references}
\end{document}